\documentclass[aps,prd,showpacs]{revtex4}
\usepackage{amsmath,amsthm,latexsym,amssymb,amsfonts,parskip,epsfig}
\allowdisplaybreaks

\newtheorem{lemm}{Lemma}[section]
\newtheorem{prop}[lemm]{Proposition}

\newcommand{\R}{\mathbb{R}}                  
\newcommand{\Z}{\mathbb{Z}}                  
\newcommand{\N}{\mathbb{N}}                  

\newcommand{\betr}[1]{\left\lvert #1 \right\rvert}

\newcommand{\expec}[1]{\left\langle #1 \right\rangle}

\newcommand{\rand}[1]{\mathbf{#1}}
\newcommand{\mat}[1]{\underline{\underline {#1}}} 

\newcommand{\avec}[1]{\underline{#1}} 

\newcommand{\aver}[1]{\overline{#1}}  

\newcommand{\qqquad}{\qquad\qquad}

\newcommand{\twolines}[2]{\genfrac{}{}{0pt}{}{\text{#1}}{\text{#2}}}
\newcommand{\teil}[1]{\subsubsection*{#1}}
\newcommand{\nl}{\boldsymbol{\lambda}}

\DeclareMathOperator{\one}{\mathbb{I}}

\begin{document}
\title{Wave propagation on a random lattice}
\date{29.12.2009}
\author{Hanno Sahlmann}
\affiliation{Institute for Theoretical Physics, Karlsruhe Institute of Technology}
\preprint{KA-TP-18-2009}
\pacs{04.60.Pp,04.60.Nc,63.50.-x}
\begin{abstract}
Motivated by phenomenological questions in quantum gravity, we consider the
propagation of a scalar field on a random lattice. We describe a procedure to
calculate the dispersion relation for the field by taking a limit of a periodic
lattice. We use this to calculate the lowest order coefficients of the
dispersion relation for a specific one-dimensional model. 
\end{abstract}
\maketitle
\section{Introduction}
In the present article we will explore the propagation of a scalar degree of
freedom on a one-dimensional lattice. The lattice is constant in time, but the
geometric information about the lattice -- encoded in a metric-like function --
is allowed to vary spatially.\footnote{In fact, as we will explain below, our
calculations are immediately applicable also to the reversed situation: a
space-time that is spatially homogeneous, but discrete and with fluctuating
geometry in the time direction.} In fact we are most interested in the case
where the geometry fluctuates in a stochastic way along the lattice. We call
this a \emph{random lattice}. The dynamics we consider for the scalar field is
that of a massless Klein-Gordon field, discretized on a lattice with variable
lattice spacing. Thus, in the limit of the lattice spacings going to zero, one
expects to obtain the well known continuum model. Here we are however interested
in the model with \emph{finite} lattice spacing, but for excitations of the
field that are on very large scales as compared to the average lattice spacing.
In this regime, the system should show propagating waves, governed by a
dispersion relation. The question we pose to ourselves is how this dispersion
relation depends on the microscopic parameters of the theory. This is very much
in analogy to the question of how to determine the speed of sound, and other
macroscopic 
properties of a (possibly amorphous) solid, from the microscopic properties of
its constituents. We are, however, not motivated by solid state physics, but by
quantum gravity. 

The incorporation of the quantum nature of gravity is likely to change the
structure of space-time on small length-scales. Different such structures have
been investigated, some motivated by full-fledged attempts to quantize 
gravity, some as toy models or as effective descriptions.  One paradigm is that
spacetime is fundamentally discrete, such as in the causal set approach
\cite{Bombelli:1987aa}, or in semiclassical considerations of loop quantum
gravity \cite{Gambini:1998it,Sahlmann:2002qj,Sahlmann:2002qk}. Another paradigm,
pioneered by Wheeler, has spacetime smooth, save for localized defects
\cite{Bernadotte:2006ya}. Our model here is an instance of the former, but could
maybe also be read as a 
model of space-time defects in the limit where their density becomes very high.
The model is very similar to the ones considered in loop quantum gravity
\cite{Gambini:1998it,Sahlmann:2002qj,Sahlmann:2002qk}, and references therein
(and some intermediate results of the present work were already contained in
\cite{diss}).
There, dispersion relations were also calculated, but the calculations involved
simplifications, effectively replacing a random lattice by a regular one, the
parameters of which were obtained by averaging. In the present article we go
beyond this, by obtaining exact results. 

We should point out that because we are working with a free massless field in
two dimensions, the corresponding continuum model would be symmetric under the
exchange of space and time coordinates. Therefore, our model is mathematically
equivalent to one in which spatial geometry is smooth and homogeneous, but time
is a discrete variable. For simplicity, we will work in the ``discrete space,
continuous time'' picture, but we will also give the results for the opposite
case in an appendix.  

Since we are working with random lattices, our results are stochastic in nature.
So while we said that we would like to compute the coefficients in the
dispersion relation, what we actually compute are expectation values (and in one
case the variance) of such coefficients. 

Fields on random lattices have become a valuable tool in lattice gauge theory,
starting from the pioneering works \cite{Christ:1982zq,Christ:1982ck}. We are,
however, not aware of explicit results on the dispersion relations of these
fields at finite lattice spacing. This may have to to with the fact that in this
context, the random lattices are a tool to obtain statements on the continuum
limit of the theories.  

Effects of coupling to random fields on the dispersion relation have already
been studied in detail in \cite{Klinkhamer:2003ec} and \cite{Klinkhamer:2005ys}.
In these works there is no discreteness of space-time. Rather, the random fields
are an effective description of the effects of the CPT anomaly, caused by a
nontrivial space-time topology on small length scales. While these models are
vastly more physical, and can hence be used to obtain bounds on small-scale
space-time structure, one can nevertheless compare them to our model on a
mathematical level. The results bear some intriguing similarities but also some
differences. These are discussed in a bit more detail at the end of this
article.  

It should be said that our model is rather unphysical, in that it is 1+1
dimensional, and in that the field content is not realistic at all. What we hope
to have accomplished, is to show how a calculation of phenomenologically
interesting data from a model involving discrete, randomly fluctuating
space-time can be accomplished. Our calculations here can almost certainly be
generalized to more realistic models (using, for example, the
Voronoi-construction \cite{Bombelli:2004si} to obtain a random lattice in
arbitrary dimension). They would merely be more cumbersome. 

The structure of the article is as follows: In the next section we will specify
the details of the model. Section \ref{se_disp} explains the calculation of the
dispersion relation. We end with a discussion of further prospects in section
\ref{se_disc}. Two appendices contain some of the longer calculations, and the
last appendix contains the dispersion relation for the case of continuous space
and discrete time.   

\section{The model}
We consider a bosonic field propagating on discrete 
space and continuous time. (The case of discrete time and continuous space is
briefly discussed in appendix \ref{app_timespace}.) To keep things simple, we
work in 1+1 dimensions. The field is a function on the space $\R\times\Z$. It
will be denoted by $\phi_n(t)$. The 
geometry of space is encoded in a time-independent, positive function $g_n$.
Often it is also
useful to re-express $g_n$ as $g_n=l_n^{-2}$. At this point we will make no
further assumptions on
$g_n$, but we will later assume that it is \emph{random} in a certain specific
sense. 

The action for our model reads 
\begin{equation}
 S=\int\text{d}t\sum_{n\in\Z}
\frac{1}{2}\left[\dot\phi_n^2-g_n(\partial^+\phi)_n^2 \right]. 
\end{equation}
Here $\partial^+\phi$ is the forward discrete derivative,
$(\partial^+\phi)_n=\phi_{n+1}-\phi_n$. It was used for simplicity. A more
symmetric derivative could be used in its place. 
We will also use its adjoint $\partial^-$ with respect to the sum over $\Z$. 
$S$ can be viewed as a discretization of the action for a free massless scalar
in the continuum, but ultimately an action of this form may be derived from a
theory of quantum gravity, and thus regarded as fundamental. 

The equations of motion are 
\begin{equation}
\label{eq_eom}
 \ddot\phi_n+\Delta_g\phi_n\equiv \ddot\phi_n-g_n\phi_{n+1}-g_{n-1}\phi_{n-1}
+(g_{n-1}+g_{n+1})\phi_n=0
\end{equation}
where we have introduced the discrete, positive definite Laplacian 
\begin{equation}
 \Delta_g=\partial^- g_n\partial^+.
\end{equation}
For completeness, we also state the Hamiltonian 
\begin{equation}
 H=\frac{1}{2}\sum_n \pi_n^2+g_n(\partial^+\phi)^2_n. 
\end{equation}
It is of the form that is obtained under certain assumptions in loop quantum
gravity \cite{Sahlmann:2002qj,Sahlmann:2002qk}. 

The form of the function $g_n$ is obviously vital for the definition of the
model. For our consideration, three cases are of interest: 
\begin{enumerate}
 \item $g_n$ is constant
 \item $g_n$ is periodic, with a certain period $N$ in $\Z$. 
 \item $g_n$ is obtained as an instance of a random process. 
\end{enumerate}
Case 3 is the one we want to consider. The simplest situation would be that the
value of $g_n$, would, for any given $n$, be determined by the sampling of a
certain random variable $\rand{g}$. Another way to state this is that there 
are independent, equally distributed random variables $\rand{g}_n$ and the
function $g_n$ is a sampling of these. More complicated situations (for example
correlations between the $\rand{g}_n$) are also conceivable. 

As we will see, it will be important to have information about the distribution
of values $\{g_n\}$ in the sampling of $\{\rand{g}_n\}$. Of particular
importance for us will be the \emph{mean} of the $l_n$, as well as some related
quantities. Such information may in principle be obtained through the law of
large numbers, or a central limit theorem.

To be concrete let us fully specify one simple model: To make the description
simple, we will not specify the distribution of the variables $\rand{g}_n$, but
those of the $\rand{l}_n$. We denote expectation values by angular brackets
$\expec{\cdot}$. Let all the random variables $\rand{l}_n$ be independent
Gaussian distributions with first and second moments
\begin{equation}
\label{gaussian}
\expec{\rand{l}_n}=l,\qquad \expec{(\rand{l}_n-l)^2}=d^2.
\end{equation}

Returning to the list of cases from above, the first case is that of an
equidistant lattice. It is immediately solvable. The second case is still
solvable and we will use it in order to analyze the case we are really
interested in. We will see that we can obtain the lowest order terms of the
dispersion relation for case 3 in a limit of case 2.
This will be explained in the following section.  
\section{Dispersion relations}
\label{se_disp}
The problem of wave propagation on a random lattice is certainly a complicated
one. In fact, we will see indications that it is not even always well defined,
i.e.\ that there is not always a long wavelength limit in which something
resembling plane waves propagates on the lattice. The question is how to
identify the cases in which the problem is well defined, and how to extract
characteristic long wavelength quantities. We will not address these problems 
in all generality. But we will obtain a  formula that, in a certain limit, gives
the first few terms in the dispersion relation for the field. This limit is by
no means well defined for all random lattices. If it is ill-defined, this is a
strong hint that there is no regime in which the lattice supports propagating
waves. 

\begin{figure}
\centerline{\epsfig{file=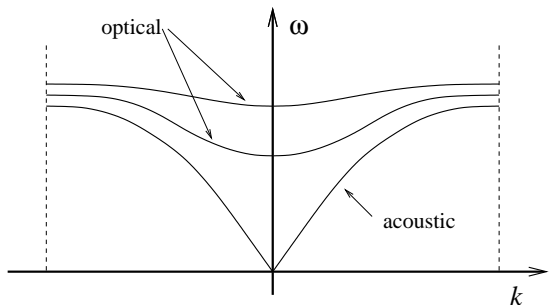, height=4cm}}
\caption{\label{fi3.1}Optical and acoustic branches in the dispersion relation}
\end{figure}

To start our quantitative discussion, we consider the case of a regular lattice.
If all the $l_i$ are equal (to $l$, say), it is easy to solve the equations of
motion of \eqref{eq_eom}. The solutions are ``plane waves''
\begin{equation}
\label{eq3.10}
\phi_{n}(t,k)=e^{i(kln-\omega(k)t)},
\qquad
\omega^2(k)=\frac{2}{l^2}\left(1-\cos(kl)\right)= k^2
-\frac{l^2}{12}k^4+O(k^6).
\end{equation}
For the general case on the other hand, with generically all $l_i$
different, it is not possible to explicitly write down any solution
to the equations of motion.
The analysis we are aiming at in the present section 
lies somewhere in-between
these two extreme cases. We will make an assumption on the $l_i$ 
under which we are able to treat the system analytically and try 
to remove it at the end of the analysis: 
Let us assume that the system is periodic with $N\in\N$ the
length of period. More precisely we assume that $g_{n+N}=g_n$ for all
$n\in\Z$. We introduce the notation $\phi_n^{(z)}\doteq\phi_{n+zN}$
with $n\in\{0,1,\ldots,N-1\}$
and make the Ansatz
\begin{equation}
\label{eq3.2}
  \phi_n^{(z)}(t)=c_n\exp i(Lzk-\omega t), \qquad L=\sum_{n=0}^{N-1} l_n.
\end{equation}
This Ansatz turns the equations of motion \eqref{eq_eom}
into an eigenvalue problem for $\avec{c}$ and $\omega$: \eqref{eq3.2}  
is a solution iff 
\begin{equation}
\label{eq3.23} 
  \mat{M}\avec{c}=\omega^2\avec{c}\text{ where }
  \mat{M}=
  \begin{pmatrix}
    g_{N-1}+g_0&-g_0&0&\ldots&0&-g_{N-1}e^{ikL}\\
    -g_0&g_0+g_1&-g_1&0&\ldots&0\\
    0&-g_1&g_1+g_2&-g_2&0&\ldots\\
    \hdotsfor{6}\\
    -g_{N-1}e^{-ikL}&0&\ldots&0&-g_{N-2}&g_{N-2}+g_{N-1}
  \end{pmatrix}.
\end{equation}
The eigenvalues $\omega_{0}\ldots\omega_{N-1}$ represent the different 
branches of the dispersion relation. We will presently see that there is 
one branch, denoted $\omega_{\text{ac}}$ in the following, with  
$\omega_{\text{ac}}(k)\rightarrow 0$ for $k\rightarrow 0$. Following 
the custom of condensed matter physics, we call this branch
\textit{acoustic} in contrast to the \textit{optical} branches nonzero 
at $k=0$.\footnote{The acoustic branch accommodates arbitrarily low frequencies,
and involves the field at neighboring sites move approximately in parallel. In
solids, these modes can be excited by sound waves, hence their name. The other
branches in solids are high frequency oscillations that can be excited by
electromagnetic radiation, typically microwaves (see for example
\cite{kittel}).} The situation is sketched in figure \ref{fi3.1}.  
As we are interested in the low energy (i.e. small $\omega$) behavior
of the field, the acoustic branch is the relevant one for our purpose 
and we will compute its small $k$ behavior in the following. 
Let us start by making the Ansatz
\begin{equation}
\label{eq3.3}
  \omega_{\text{ac}}^2(k)= w(1)\betr{k}+w(2)
  k^2+w(3)\betr{k}^3+\ldots,  
\end{equation}
explicitly forcing $\omega_{\text{ac}}(0)$ to be zero. Accordingly, we 
expand $\det(\mat{M}-\omega^2\one)$:
\begin{equation}
\label{eq3.4}
\det(\mat{M}-\omega^2\one)=\sum_{i=0}^{N-1}\omega^{2i}\sum_{j=0}^\infty
w(i,j) \betr{k}^j . 
\end{equation}
We want to determine the coefficients $w(i)$ of \eqref{eq3.3}. 
The coefficients $w(i,j)$ of the expansion \eqref{eq3.4} on the other hand can
be explicitly calculated.
The calculation is cumbersome, so we have relegated it to 
appendix \ref{app_det}. We find
\begin{prop}
\label{pro_det}
For $\mat{M}$ of the form \eqref{eq3.23},
  \begin{equation}
  \begin{split}
    \det(\mat{M}-\omega^2\one)=&-2g_0\ldots g_{N-1}(1-\cos L k)\\
    &+\omega^2 Ng_0\ldots g_{N-1}\sum_{i=0}^{N-1} g_i^{-1}\\
    &+\omega^4 g_0\ldots g_{N-1}\sum_{0\leq i<j\leq
      N-1}(j-i)[N-(j-i)]g_i^{-1}g_j^{-1}\\
    &+O(\omega^6).
    \end{split}
  \end{equation}
\end{prop}
Froom this, the  coefficients $w(i,j)$ can be read off. By solving the  
eigenvalue equation $\det(\mat{M}-\omega^2\one)$ order by order, we then obtain
the low order coefficients of the dispersion relation \eqref{eq3.3}. 
We use the shorthands $c_{ij}\doteq (j-i)[N-(j-i)]$ and 
\begin{equation}
\aver{f}\doteq \frac{1}{N}\sum_{i=0}^{N-1}f_n  
\end{equation}
for the average of some quantity over the period of the lattice. Then we have
\begin{equation}
\label{eq3.12}
\begin{split}
  \omega_{\text{ac}}^2(k)&=
  \frac{L^2}{N^2}\frac{1}{\aver{g^{-1}}}k^2
  +\left(\frac{L^4}{N^6}\frac{\sum_{i<j}c_{ij}g_i^{-1}g_j^{-1}}
      {(\aver{g^{-1}})^3}-\frac{1}{12N^2}\frac{L^4}{\aver{g^{-1}}{}}
\right){k}^4+O\left({k}^6\right)\\
&=
  \frac{\aver{l}^2}{\aver{l^2}}\betr{k}^2
  +\left(\frac{1}{L^2}\frac{\aver{l}^6}{(\aver{l^2})^3}
    \sum_{i<j}c_{ij}l_i^2l_j^2
    -\frac{L^2}{12}\frac{\aver{l}^2}{\aver{l^2}}
  \right)\betr{k}^4+O\left(\betr{k}^6\right). 
\end{split}
\end{equation}
This is a remarkable formula, and one of the main results of the present work.
It is an \emph{exact} result for the 
lowest orders of the dispersion relation of the field $\phi$ propagating on a
periodic lattice. 
As such, it reproduces the elementary result \eqref{eq3.10} for the case of the
regular lattice, by setting 
$l_n=l$ for all $n$ and using 
\begin{equation}
\label{eq3.11}
  \sum_{0\leq i<j\leq N-1}c_{ij}=\frac{1}{12}N^2(N^2-1).
\end{equation}
We note that the square of the velocity of propagation, 
\begin{equation}
\label{eq_speed}
 c^2:= \frac{\aver{l}^2}{\aver{l^2}}
\end{equation}
is neatly expressed in terms of averages over lattice spacings. The formula for
the next order coefficient, 
\begin{equation}
\label{eq_ell}
\ell^2:=\frac{1}{L^2}\frac{\aver{l}^6}{(\aver{l^2})^3}
    \sum_{i<j}c_{ij}l_i^2l_j^2
    -\frac{L^2}{12}\frac{\aver{l}^2}{\aver{l^2}}
\end{equation}
is certainly more complicated, but also given in terms of such averages. This is
more than just aesthetically pleasing: It suggests that \eqref{eq3.12} may
survive the large-$N$ limit. Moreover, that when interpreting the 
$l_i$ as given by a sampling of some random process, such averages may be
expressible by averages under this process, through some ergodic-type results.

Let us come back to our original motivation: We were interested in the case of a
periodic lattice, because, for 
the $l_n$ determined by some random process, the limit $N\rightarrow\infty$,
with the average length 
$\expec{\rand{l}_i}=l$ held fixed, gives an infinite random lattice. Thus we
would like to consider the dispersion relation \eqref{eq3.12} in that limit. We
see, however, no possibility to discuss this limit without fixing a distribution
for the $l_i$. Central limit theorems would make statements about the
expectation value of the mean $\aver{\rand{l}}$ and functions of it in the limit
$N\rightarrow\infty$ for large classes of distributions. They can similarly be
used to obtain statements about the mean of the squares, $\aver{\rand{l}^2}$, in
this limit. But in our case, we need to investigate 
a function of both, $\aver{l}$ and $\aver{l^2}$ in the case of the first term in
the expansion \eqref{eq3.12}. The second term in \eqref{eq3.12} additionally
depends on $N$ and on a curious linear combination of $l^2_il^2_j$, thus 
further complicating the situation. Let us therefore consider the situation
described in \eqref{gaussian} in which the $l_i$ are independently Gaussian
distributed, with average $\expec{l_i}=l$ and variance
$\aver{\rand{l}_i^2}-\aver{\rand{l}_i}^2=d^2$. What we want to do is to compute
expectation values of the coefficients \eqref{eq3.12} under this distribution,
in the limit $N\rightarrow\infty$.  

Let us first consider the coefficient $c^2$ of the $|k|^2$-term in
\eqref{eq3.12}. 
In appendix \ref{app_gauss} we show that 
while we can not compute its expectation value directly, we can expand it in
powers of $d/l$, compute the expectation values of the first few terms, and take
$N$ to infinity. The result 
\begin{equation}
\lim_{N\rightarrow\infty}\expec{\frac{\aver{\rand{l}}^2}{\aver{\rand{l}^2}}}
=1-\frac{d^2}{l^2}+\frac{d^4}{l^4}-\frac{d^6}{l^6}+\ldots
\end{equation}
strongly suggests, that 
\begin{equation}
\label{eq_c}
\lim_{N\rightarrow\infty}\expec{\frac{\aver{\rand{l}}^2}{\aver{\rand{l}^2}}}=
\frac{1}{1+\frac{d^2}{l^2}}\equiv 
\frac{\expec{\aver{\rand{l}}^2}}{\expec{\aver{\rand{l}^2}}}.
\end{equation}
\begin{figure}
\centerline{\epsfig{file=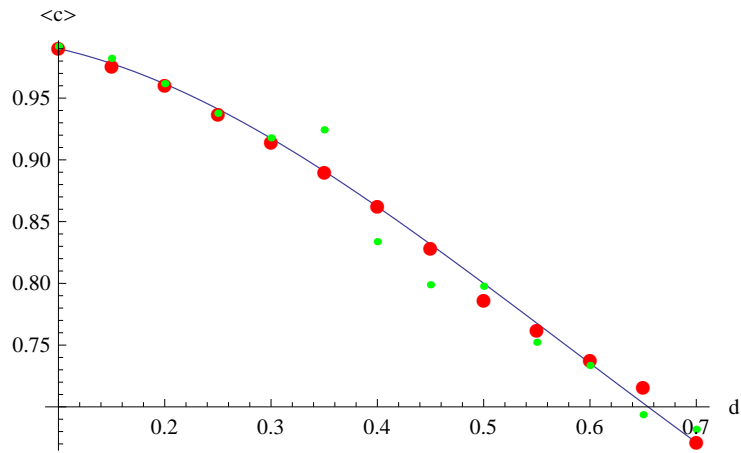, height=6cm}}
\caption{\label{fi_mc1} Theory and numerical simulation for the coefficient
$c^2$: For $l=1$, we plot the theoretical result \eqref{eq_c} as a function of
$d$ (continuous curve) together with two numerical simulations ($N=100$: small
dots, $N=1000$: large dots). Convergence is nicely visible.}
\end{figure}
For independent confirmation we have also checked this result on a computer, by
calculating instances of $c^2$ for randomly generated sets $\{l_i\}$. Formula
\eqref{eq_c} describes the numerical results very well. An example is given in
figure \ref{fi_mc1}.  

Furthermore, we are able to calculate the variance of $c^2$, again as a series
in powers of $d/l$, and take the $N\rightarrow\infty$ limit term-wise. It turns
out that the variance \emph{vanishes} in this limit:
\begin{equation}
\lim_{N\rightarrow\infty}\expec{\rand{c}^4}-
\lim_{N\rightarrow\infty}\expec{\rand{c}^2}^2=0.
\end{equation}Details are again given in appendix \ref{app_gauss}.
  
\begin{figure}
\centerline{\epsfig{file=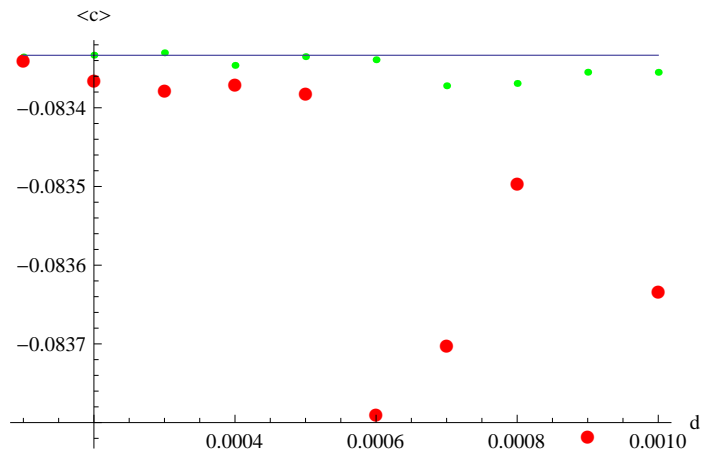, height=6cm}}
\caption{\label{fi_mc2} Theory and numerical simulation for the coefficient
$\ell^2$: For $l=1$, we plot the theoretical result \eqref{eq_C} as a function
of $d$ (continuous curve) together with two numerical simulations ($N=100$:
small dots, $N=1000$: large dots). Correspondence between theory and numerics is
visible for very small values of $d$ only, and convergence apparently gets worse
for larger $N$.}
\end{figure}

The situation is more complicated with the term proportional to $|k|^4$.
Expansion in $d/l$ (see appendix) gives  
\begin{equation}
\expec{\rand{\ell^2}}=-\frac{l^2}{12}+\frac{l^2}{12
N}(-4N^2+N-6)\left(\frac{d}{l}\right)^2+O\left(\left(\frac{d}{l}\right)^3\right)
\end{equation}
which does not converge term by term in the limit $N\rightarrow\infty$.  But
this does not say that the limit does not exist. In fact, if one would
extrapolate from the result for the coefficient $c^2$,
\begin{equation}
\label{eq_C}
\begin{split}
\lim_{N\rightarrow\infty}\expec{\rand{\ell^2}}&=\lim_{N\rightarrow\infty}\expec{
\frac{1}{\rand{L}^2}\frac{\aver{\rand{l}}^6}{(\aver{\rand{l}^2})^3}   
\sum_{i<j}c_{ij}\rand{l}_i^2\rand{l}_j^2}-\lim_{N\rightarrow\infty}\expec{\frac{
\rand{L}^2}{12}\frac{\aver{\rand{l}}^2}{\aver{\rand{l}^2}}}\\
&\overset{?!}{=}
\frac{1}{N^2}\frac{\expec{\aver{\rand{l}}}^4}{\expec{\aver{\rand{l}^2}}^3}\sum_{
i<j}c_{ij}
\expec{\rand{l}_i^2\rand{l}_j^2}-\frac{N^2}{12}
\frac{\expec{\aver{\rand{l}}}^4}{\expec{\aver{\rand{l}^2}}}\\
&=\frac{1}{12}\frac{l^4}{l^2+d^2}(N^2-1)-\frac{1}{12}\frac{l^4}{l^2+d^2}N^2\\
&=-\frac{1}{12}l^2\frac{1}{1+\frac{d^2}{l^2}}
\end{split}
\end{equation}
one expects the result to be finite. Numerical results for small $d/l$ indeed
point towards convergence (figure \ref{fi_mc2}), but 
for larger $d/l$, convergence in the limit $N\rightarrow\infty$ could not be
established. It may be there but too slow to be seen. 

For completeness, we have also considered the case of discrete time and
continuous space in appendix \ref{app_timespace}. The resulting dispersion
relation is
\begin{equation}
\omega^2(k)=\frac{1}{c^2}k^2-\frac{\ell^2}{c^6}k^4+\ldots.
\end{equation} 
However, we have not considered the expectation values of the coefficients in
any detail.  
\section{Discussion and outlook} 
\label{se_disc}

In the present paper we have calculated the (expectation values of the) first
coefficients in the dispersion relation 
\begin{equation}
\omega^2(k)=c^2|k|^2+\ell^2 |k|^4 +\ldots
\end{equation}
for a scalar field propagating on what one could call a Gaussian random lattice
in 1+1 dimensions. The first term, also setting the phase velocity, is unitless
in our convention, thus the natural value is 1. We have seen 
\begin{equation}
\lim_{N\rightarrow\infty}\expec{\rand{c}^2}= \frac{1}{1+\frac{d^2}{l^2}}
\end{equation}
i.e.\ it is in fact very close to 1, as long as the variance of the lattice
spacing is small. We have also presented very strong evidence that its variance
is strictly zero for an infinitely extended lattice, thus one does not even have
to take into account that in principle it has a probabilistic nature. In more
realistic models, this coefficient would be very interesting for phenomenology:
As was demonstrated in \cite{Bernadotte:2006ya}, experiments and observations
are very sensitive to differences of propagation speed of the fields in nature. 
The second term had a more complicated structure, and convergence in the desired
limit could not be demonstrated to complete satisfaction. We have conjectured
the limit 
\begin{equation}
\lim_{N\rightarrow\infty}\expec{\rand{\ell^2}}=-\frac{1}{12}\frac{l^2}{1+\frac{
d^2}{l^2}}.
\end{equation}
Its scale is set by the average lattice spacing squared, $l^2$, as long as the
variance of the lattice spacing is small. We have also briefly discussed a model
in which the roles of space and time are, switched, i.e., time is discrete and
space is continuous. There we see potential problems with causality, with phase
and group velocity increasing due to the lattice effects.    

An interesting observation was made by a referee of a draft version of this
article: The first two terms of the dispersion relation for the field on a
regular lattice \eqref{eq3.10} can be turned into the first two terms of the
dispersion relation on the random lattice, by rescaling $k\rightarrow k l$ and 
dividing by $\langle\aver{l^2}\rangle$ in \eqref{eq3.10}.  It
is very well possible, that this procedure gives the correct dispersion
relation on the random lattice to all orders. This does not detract from the
value of the results presented in the article: While there may be a shortcut, it
is still neccesary to show that the shortcut indeed provides the right results.
In fact, one big motivation for the article was that several different
``shortcuts'' were used in the literature on quantum gravity phenomenology, with
different results.  We should also point out that, despite the fact that the
dispersion relation of
the random lattice seems to be \emph{obtainable from} that of a regular lattice
by the above scaling operations, it \emph{is not} that of such a regular
lattice.

Similar, but much more detailed, calculations were carried out in
\cite{Klinkhamer:2003ec,Klinkhamer:2005ys} for random fields coupled to scalar
and electromagnetic fields in physical space-time dimensions. Since these were
continuum models, one expects differences in the results: By reformulation our
model in terms of a continuum field, one can show that in the equations of
motion, there are correction terms of arbitrary derivative order, as compared to
the equations of motion for the free scalar field. In contrast, the fields of  
\cite{Klinkhamer:2003ec,Klinkhamer:2005ys} receive only corrections in terms of
first derivatives of the field. Still the results on the dispersion relation are
similar: The phase velocity receives a correction \emph{downward}, and the
coefficient of $k^4$ is negative, thus insuring causal propagation. Moreover,
the mathematical structure is similar: In both cases, the second moment of the
random field enters the phase velocity, and 
the coefficient of $k^4$ is given by moments the autocorrelation of the random 
field, the first moment in case of  \cite{Klinkhamer:2003ec,Klinkhamer:2005ys},
the second moment (of $l_n^2$) in the present model. 

Besides finishing the discussion begun in the present work, by showing
convergence or divergence of the second coefficient $\ell^2$, and studying the
complete model for more generic distributions for the random lattice,
it would be interesting to study the nature of the eigen\textit{vectors}
$\avec{c}$, at least in low order in $k$. For example: Does the one for the
acoustic branch really look like a plane wave, at least for large $N$? Obvious
further goals are the extension of the formalism to vector and fermionic fields
and to physical dimension. At that point, one would be in a position to compare
the models with experiment and put bounds on parameters of, for example, loop
quantum gravity. 

\section*{Acknowledgments}
We thank F.R.\ Klinkhamer for sharing his insights about the 
phenomenology of quantum gravity, partially motivating the present work, and for
numerous helpful comments on an earlier version of this article.  This work has been partially supported by the Spanish MICINN project No. FIS2008-06078-C03-03. 

\appendix
\section{Proof of proposition \ref{pro_det}}
\label{app_det}
It is an elementary combinatorial fact that
  \begin{equation}
    \label{eq3.24}
    \det(\mat{M}-\omega^2\one)=\sum_{i=0}^{N-1} (-1)^i\omega^{2i}
    \left(\twolines{sum over all $(N-i)\times (N-i)$}
        {principal sub determinants of $\mat{M}$}\right).
  \end{equation}
Therefore the proof of the theorem reduces to the 
calculation of numerous sub-determinants of $\mat{M}$. 
These calculations are tedious but elementary. As a preparation, we
observe that for $n$ in $\{0,1,\ldots,N-1\}$
\begin{equation}
\label{eq3.6}
\begin{vmatrix}
    g_0+g_1&-g_1&&&0\\
    -g_1&g_1+g_2&-g_2&&\\
    &\ddots&\ddots&\ddots&\\
    &&&&-g_n\\
    0&&&-g_n&g_n  
\end{vmatrix}=
g_0\ldots g_n
\end{equation}
by repeatedly adding all other columns to the first one and pulling
out factors. Repeated use of the linearity of the determinant in the last
column of \eqref{eq3.6} yields
\begin{equation}
\label{eq3.7}
\begin{vmatrix}
    g_0+g_1&-g_1&&&0\\
    -g_1&g_1+g_2&-g_2&&\\
    &\ddots&\ddots&\ddots&\\
    &&&&-g_{n-1}\\
    0&&&-g_{n-1}&g_{n-1}+g_{n}  
\end{vmatrix}=g_0\ldots g_{n}\sum_{i=0}^{n} g_i^{-1},
\end{equation}
another identity which will be used frequently. Finally we introduce
the abbreviation $q\doteq\exp(-ik)$.\newline
We turn now to the calculation of the lowest order coefficients in
\eqref{eq3.24}.  
\teil{Calculation of $\det\mat{M}$:}
\begin{align*}
  \det \mat{M}&= 
  \begin{vmatrix}
    g_{N-1}(2-q-q^{-1})&&&&&g_{N-1}(1-q^{-1})\\
    &g_0+g_1&-g_1&&&\\
    &-g_1&g_1+g_2&-g_2&&\\
    &&\ddots&\ddots&\ddots&\\
    &&&&&-g_{N-2}\\
    g_{N-1}(1-q)&&&&-g_{N-2}&g_{N-2}+g_{N-1}
  \end{vmatrix}\\
  \intertext{by adding all columns to the first column and subsequently
    all rows to the first one,} 
  &=g_{N-1}(2-q-q^{-1})
  \begin{vmatrix}
    g_0+g_1&-g_1&&&0\\
    -g_1&g_1+g_2&-g_2&&\\
    &\ddots&\ddots&\ddots&\\
    &&&&-g_{N-2}\\
    0&&&-g_{N-2}&g_{N-2}
  \end{vmatrix}
  \intertext{by pulling out a factor and eliminating the entry in the
    upper right and lower left corners,}
  &=g_{N-1}(2-q-q^{-1})g_0
\begin{vmatrix}
    g_1+g_2&-g_2&&&0\\
    -g_2&g_2+g_3&-g_3&&\\
    &\ddots&\ddots&\ddots&\\
    &&&&-g_{N-2}\\
    0&&&-g_{N-2}&g_{N-2}
  \end{vmatrix}\\
\intertext{by adding all columns to the first column and subsequently
    all rows to the first one and pulling out a factor,}
&=g_{0}\ldots g_{N-1}(2-q-q^{-1})
\end{align*}
by applying \eqref{eq3.6}.  
\teil{Calculation of the $(N-1)\times(N-1)$ sub-determinants:}
Let $0<n<N-1$. We consider computing the sub-determinant of $\mat{M}$ 
where row and column $n$ are deleted. 
\begin{align*}
  \det\mat{M}'_n&=
  \begin{vmatrix}
    g_{N-1}+g_{0}&-g_0&&&-g_{N-1}q^{-1}\\
    \ddots&\ddots&\ddots&&\\
          &g_{n-1}+g_n&0     &&\\
          &0     &1     &0&\\
          &      &0     &g_{n+1}+g_{n+2}&\\  
          &      &\ddots&\ddots&\ddots\\
    -g_{N-1}q&&&-g_{N-2}&g_{N-2}+g_{N-1}
  \end{vmatrix}\\
  &=\begin{vmatrix}
    g_{N-1}+g_{0}&-g_0&\\
    -g_0&&\\
    &\ddots&\\
    &&-g_{n-1}\\
   &-g_{n-1}&g_{n-1}+g_n
 \end{vmatrix}
 \begin{vmatrix}
    g_{n+1}+g_{n+2}&-g_{n+2}&\\
    -g_{n+2}&&\\
    &\ddots&\\
    &&-g_{N-2}\\
    &-g_{N-2}&g_{N-2}+g_{N-1}
 \end{vmatrix}\\
&\qquad -q^{-1}g_{N-1}
  \begin{vmatrix}
    g_{N-1}+g_{0}&-g_0&&&1\\
    -g_0&\ddots&&&\\
          &     &1     &&\\
          &      &&\ddots&0\\
    0&&&-g_{N-2}&0
  \end{vmatrix}
-q g_{N-1}
  \begin{vmatrix}
    0&-g_0&&&0\\
    0&\ddots&&&\\
          &     &1     &&\\
          &      &&\ddots&-g_{N-2}\\
    1&&&0&0
  \end{vmatrix}\\
&\qquad -g^2_{N-1}
\begin{vmatrix}
    g_0+g_1&-g_1&&&0\\
    -g_1&\ddots&&&\\
    &&1&&\\
    &&&\ddots&-g_{N-3}\\
    0&&&-g_{N-3}&g_{N-3}+g_{N-2}
\end{vmatrix}\\
\intertext{by expanding in the first and last column of the
  matrix. It is not hard to see that the determinants in the terms
  proportional to $q$ and $q^{-1}$ vanish: The corresponding matrices
  can be brought to a form where they contain a zero column by simple
  column operations. The remaining determinants can be treated using
  \eqref{eq3.7}:}
&=g_{N-1}g_0\ldots g_n\left(g_{N-1}^{-1}+\sum_{i=0}^{n}g_i^{-1}\right)
g_{n+1}\ldots g_{N-1}\left(\sum_{i=n+1}^{N-1}g_i^{-1}\right)\\
&\qquad
-g_{N-1}^2g_0\ldots g_n\left(\sum_{i=0}^{n}g_i^{-1}\right)
g_{n+1}\ldots g_{N-2}\left(\sum_{i=n+1}^{N-2}g_i^{-1}\right)\\
&=g_0\ldots g_{N-1}\sum_{i=0}^{N-1}g_i^{-1}
\end{align*}
The cases where $n=0$ and $n=N-1$ have to be treated separately,
either by an explicit calculation or by appealing to the symmetry of
the problem under cyclic permutations of $g_0\ldots g_{N-1}$.
They yield the same result.
\teil{Calculation of the $(N-2)\times(N-2)$ sub-determinants:}
The calculation of the $(N-2)\times(N-2)$ sub-determinants proceeds
analogously to that in the last paragraph.   
Let $0<n<m<N-1$ and consider the sub-determinant of $\mat{M}$ 
where row and column $m$ and $n$ are deleted. Again we start by 
expanding linearly in the first and the last column:
\begin{align*}
  &\det\mat{M}'_{nm}=\\
  &\qqquad\begin{vmatrix}
    g_{N-1}+g_{0}&-g_0&\\
    -g_0&&\\
    &\ddots&\\
    &&-g_{n-1}\\
   &-g_{n-1}&g_{n-1}+g_n
 \end{vmatrix}
 \begin{vmatrix}
    g_{n+1}+g_{n+2}&-g_{n+2}&\\
    -g_{n+2}&&\\
    &\ddots&\\
    &&-g_{N-2}\\
    &-g_{m-1}&g_{m-1}+g_{m}
 \end{vmatrix}\\
&\qqquad\qqquad\cdot
 \begin{vmatrix}
    g_{m+1}+g_{m+2}&-g_{m+2}&\\
    -g_{n+2}&&\\
    &\ddots&\\
    &&-g_{N-2}\\
    &-g_{N-2}&g_{N-2}+g_{N-1}
 \end{vmatrix}\\
&-q^{-1}g_{N-1}
  \begin{vmatrix}
    g_{N-1}+g_{0}&-g_0&&&&&1\\
    -g_0&\ddots&&&&&\\
          &&1&&&&\\
          &&&\ddots&&&\\
          &&&&1&&\\
          &&&&&\ddots&0\\
    0&&&&&-g_{N-2}&0
  \end{vmatrix}
-q g_{N-1}
  \begin{vmatrix}
    0&-g_0&&&&&0\\
    0&\ddots&&&&&\\
       &&1&&&&\\
          &&&\ddots&&&\\
          &&&&1&&\\
          &&&&&\ddots&-g_{N-2}\\
    1&&&&&0&0
  \end{vmatrix}\\
&\qqquad -g^2_{N-1}
\begin{vmatrix}
    g_0+g_1&-g_1&&&&&0\\
    -g_1&\ddots&&&&&\\
       &&1&&&&\\
          &&&\ddots&&&\\
          &&&&1&&\\
    &&&&&\ddots&-g_{N-3}\\
    0&&&&&-g_{N-3}&g_{N-3}+g_{N-2}
\end{vmatrix}\\
\intertext{Again it is not hard to see that the determinants in the 
  terms proportional to $q$ and $q^{-1}$ vanish. By means of
  \eqref{eq3.7} we get}
&\qquad=g_{N-1}g_0\ldots g_{n}
\left(g_{N-1}^{-1}+\sum_{i=0}^{n}g_i^{-1}\right)
g_{n+1}\ldots g_{m}\left(\sum_{i=n+1}^{m}g_i^{-1}\right)
g_{m+1}\ldots g_{N-1}\left(\sum_{i=m+1}^{N-1}g_i^{-1}\right)\\
&\qquad\quad-g^2_{N-1} g_0\ldots g_{n}
\left(\sum_{i=0}^{n}g_i^{-1}\right)
g_{n+1}\ldots g_{m}\left(\sum_{i=n+1}^{m}g_i^{-1}\right)
g_{m+1}\ldots g_{N-2}\left(\sum_{i=m+1}^{N-2}g_i^{-1}\right).
\end{align*}
Again, the case where the first or the last row and column get deleted 
have to be considered in a separate calculation or treated by symmetry 
arguments. The same result is obtained in these cases.\\ 
When summing over all such determinants, by summing over pairs $(m,n)$, the
result further simplifies:
\begin{equation}
\sum_{0\leq m<n\leq N-1} \det\mat{M}'_{nm} =g_0\ldots g_{N-1}\sum_{0\leq i<j\leq
N-1}
(j-i)[N-(j-i)] g_i^{-1}g_j^{-1}.
\end{equation}
\section{Calculation of expectation values}
\label{app_gauss}
In the present appendix we collect some calculations regarding expectation
values 
of the coefficients of the dispersion relation. 

To make the notation more simple, we will not distinguish
between the random variables $\{\rand{l}_i\}$ and a sample $\{l_i\}$ of the
random process anymore. The context will hopefully make the distinction clear. 

We start with the lowest order coefficient. We want to calculate
\begin{equation}
 \expec{\rand{c}^2}\equiv \expec{\frac{\aver{\rand{l}}^2}{\aver{\rand{l}^2}}}. 
\end{equation}
It is convenient to go over to new random variables 
\begin{equation}
\nl_n:= \frac{\rand{l}_n-l}{d}. 
\end{equation}
These are Gaussian distributed with expectation zero and spread one. Also
introducing 
\begin{equation}
\rand{c}_1:=\sum_{n=0}^{n-1}\nl_i, \qquad \rand{c}_2=\sum_{n=0}^{n-1}\nl^2_i,
\qquad \delta=\frac{d}{l}
\end{equation}
we can rewrite the expectation value as 
\begin{equation}
\expec{\frac{\aver{\rand{l}}^2}{\aver{\rand{l}^2}}}=
\expec{\frac{\rand{c}_1^2d^2+2Nl\rand{c}_1d+N^2l^2}{N\rand{c}_2
d^2+2Nl\rand{c}_1d+N^2l^2}}=
\expec{\frac{\rand{c}_1^2\delta^2+2N\rand{c}_1\delta+N^2}{N\rand{c}
_2\delta^2+2N\rand{c}_1\delta+N^2}}.
\end{equation}
We were not able to compute this expectation value exactly. We can however
expand it in a series that converges for small variance $\delta$. To that end,
we Taylor-expand in $\delta$, obtaining 
\begin{equation}
\begin{split}
\frac{c_1^2\delta^2+2Nc_1\delta+N^2}{Nc_2\delta^2+2Nc_1\delta+N^2}=
1
+&\left(c_1^2-c_2 N\right)\left(\frac{\delta}{N}\right)^2
-2\left(c_1^3-c_1 c_2 N\right)\left(\frac{\delta}{N}\right)^3\\
&+\left(4 c_1^4-5 c_1^2 c_2 N+c_2^2 N^2\right)\left(\frac{\delta}{N}\right)^4\\
&-4\left(2 c_1^5-3 c_1^3 c_2
   N+c_1 c_2^2 N^2\right)\left(\frac{\delta}{N}\right)^5\\
&+\left(16 c_1^6-28 c_1^4 c_2 N+13 c_1^2 c_2^2
   N^2-c_2^3 N^3\right)\left(\frac{\delta}{N}\right)^6+O\left(\delta^7\right).
\end{split}
\end{equation}
The terms in this series are such that their expectation values can be computed
through tedious but 
straightforward calculations. We use 
\begin{equation}
\expec{(\nl_i)^k}=\begin{cases}0&\text{ if $k$ odd }\\
\frac{1}{\sqrt{\pi}} 2^{\frac{k}{2}}\Gamma\left(\frac{k+1}{2}\right)&\text{ if
$k$ even }
\end{cases}, \qquad \expec{\nl_i\nl_j}=0 \text{ for } i\neq j  
\end{equation}
to compute 
\begin{equation}
\begin{split}
\expec{\rand{c}_1^2}&=N\\
\expec{\rand{c}_2}&=N\\
\expec{\rand{c}_1^4}&=3N^2\\
\expec{\rand{c}_1^2\rand{c}_2}&=N(N+2)\\
\expec{\rand{c}_2^2}&=N(N+2)\\
\expec{\rand{c}_1^6}&=15N^3\\
\expec{\rand{c}_1^4\rand{c}_2}&=3N(N^2+N+3)\\
\expec{\rand{c}_1^2\rand{c}_2^2}&=N(N^2+6N+8)\\
\expec{\rand{c}_2^3}&=N(N^2+6N+8)
\end{split}
\end{equation}
Monomials in $\rand{c}_1,\rand{c}_2$ that contain odd powers of the $\nl_i$
vanish. Taking the expectation value of the 
above Taylor series then gives 
\begin{equation}
\begin{split}
\expec{\rand{c}^2}\equiv\expec{\frac{\rand{c}_1^2\delta^2+2N\rand{c}_1\delta+N^2
}{N\rand{c}_2\delta^2+2N\rand{c}_1\delta+N^2}}=
1&+(N-N^2)\left(\frac{\delta}{N}\right)^2
+(N^4-3N^3+2N^2)\left(\frac{\delta}{N}\right)^4\\
&+(-N^6+7 N^5-14 N^4+260 N^3-252 N^2)\left(\frac{\delta}{N}\right)^6+O(\delta^8)
\end{split}
\end{equation}
A remarkable aspect of this series expansion of the expectation value is that
term by term, the limit $N\rightarrow\infty$ is well defined, and suggests
\begin{equation}
\lim_{N\rightarrow\infty}\expec{\frac{\aver{\rand{l}}^2}{\aver{\rand{l}^2}}}
=1-\frac{d^2}{l^2}+\frac{d^4}{l^4}-\frac{d^6}{l^6}+\ldots
=\frac{1}{1+\frac{d^2}{l^2}}.
\end{equation}
While the above calculations do not constitute a proof in the strict sense, we
are quite confident that it is correct, since it is backed up both by heuristics
and by numerical simulations. This is discussed in the main text. 

For this term, we can even easily compute the variance with the same methods. We
first have to expand
\begin{equation}
\begin{split}
\left(\frac{c_1^2\delta^2+2Nc_1\delta+N^2}{Nc_2\delta^2+2Nc_1\delta+N^2}\right)
=1&+2\left(c_1^2-c_2 N\right)\left(\frac{\delta}{N}\right)^2\\
&+4 \left(c_1 c_2 N-c_1^3\right)\left(\frac{\delta}{N}\right)^3\\
&+3 \left(c_2^2 N^2-4 c_2 c_1^2 N+3
c_1^4\right)\left(\frac{\delta}{N}\right)^4\\
&-4 \left(3c_2^2 c_1 N^2-8 c_2 c_1^3 N+5
c_1^5\right)\left(\frac{\delta}{N}\right)^5\\
&+\left(-4 c_2^3 N^3+42 c_2^2 c_1^2 N^2-82 c_2 c_1^4 N+44
   c_1^6\right)\left(\frac{\delta}{N}\right)^6+O\left(\delta ^7\right)
\end{split}
\end{equation}
again, the expectation value of each of the terms can be evaluated and the limit
$N\rightarrow\infty$ taken. We will only give the result:
\begin{equation}
\lim_{N\rightarrow\infty}\expec{\left(\frac{\aver{\rand{l}}^2}{\aver{\rand{l}^2}
}\right)^2}=
1-2\frac{d^2}{l^2}+3\frac{d^4}{l^4}-4\frac{d^6}{l^6}+\ldots
=\frac{1}{\left(1+\frac{d^2}{l^2}\right)^2}.
\end{equation}
Therefore we obtain the very simple result that 
\begin{equation}
\lim_{N\rightarrow\infty}\expec{\rand{c}^4}-
\lim_{N\rightarrow\infty}\expec{\rand{c}^2}^2=0.
\end{equation}
The situation for the coefficient $\ell^2 $ of $\betr{k}^4$ is more murky. Again
we change variables and obtain, 
employing the same notation as above, 
\begin{equation}
\begin{split}
\rand{\ell^2}=\frac{1}{\rand{L}^2}\frac{\aver{\rand{l}}^6}{(\aver{\rand{l}^2})^3
}
    \sum_{i<j}c_{ij}\rand{l}_i^2\rand{l}_j^2
    -\frac{\rand{L}^2}{12}\frac{\aver{\rand{l}}^2}{\aver{\rand{l}^2}}=&
\frac{l^2}{N^3}\frac{(\rand{c}_1\delta+N)^4}{(\rand{c}_2\delta^2+2\rand{c}
_1\delta+N)^3}\sum_{i<j}
c_{ij}(\nl^2_i\delta^2+2\nl_i\delta+1)(\nl^2_j\delta^2+2\nl_j\delta+1)\\
&\qquad-\frac{1}{12}\frac{l^2}{N}
\frac{(\rand{c}_1\delta+N)^4}{\rand{c}_2\delta^2+2\rand{c}_1\delta+N}.
\end{split}
\end{equation}
Again, we try an expansion in $\delta$. We use
\begin{equation}
\frac{(c_1 \delta +N)^4}{\left(2 c_1 \delta
   +c_2 \delta^2+N\right)^3} =N -2c_1 \delta +\frac{3 \left(2 c_1^2-c2
N\right)}{N} \delta^2 +O(\delta^3) 
\end{equation}
and 
\begin{equation}
 \frac{(c_1\delta+N)^4}{c_2\delta^2+2c_1\delta+N}=n^3 +2 c_1
   n^2 \delta + \left(2 c_1^2 n-c_2 n^2\right)\delta ^2 
\end{equation}
to find after some calculation
\begin{equation}
\expec{\rand{\ell^2}}=-\frac{l^2}{12}+\frac{l^2}{12
N}(-4N^2+N-6)\delta^2+O(\delta^3)
\end{equation}
The problem with this expansion is that, at least term by term, the
$N\rightarrow\infty$ limit can not be taken: The term proportional to $\delta^2$
diverges linearly with $N$. This is despite some cancellation between the two
terms that comprise $C$. We will discuss the implications of this in the main
text. 
\section{Discrete time, continuous space}
\label{app_timespace}
As metioned in the introduction, we can also apply our calculations to the
situation in which time is discrete and space is continuous and homogeneous, by
exchanging the role of space and time in all formulas. 
The action in this case would read
\begin{equation*}
 S=\int\text{d}x\sum_{n\in\Z} \left[g_n(\partial^+\phi)_n-(\phi'_n)^2 \right] 
\end{equation*}
with the prime denoting the spatial derivative of the field $\phi_n(x)$.
Determination of the equations of motion proceeds in parallel with the main
text. The wave-ansatz is now
\begin{equation}
  \phi_n^{(z)}(x)=c_n\exp i(Lz\omega-k x), \qquad L=\sum_{n=0}^{N-1} l_n.
\end{equation}
The calculation of the main text then gives the dispersion relation 
\begin{equation}
  k^2(\omega)=c^2\omega^2+\ell^2\omega^4+\ldots, 
\end{equation}
with $c^2$ and $\ell^2$ exactly as in \eqref{eq_speed}, and \eqref{eq_ell}. 
We can invert this series to get to the more familiar form 
\begin{equation}
\omega^2(k)=\frac{1}{c^2}k^2-\frac{\ell^2}{c^6}k^4+\ldots.
\end{equation} 
We have not analyzed the expectation value $\expec{\rand{c}^{-2}}$ in detail,
but one can argue that since the distribution of $\rand{c}^2$ is sharp in the
limit $N\rightarrow\infty$ one should have  
\begin{equation}  
\lim_{N\rightarrow\infty}\expec{\rand{c}^{-2}}=\left(\lim_{N\rightarrow\infty}
\expec{\rand{c}^2}\right)^{-1}=\left(1+\frac{d^2}{l^2}\right)^2,
\end{equation}
i.e. we obtain a positive correction to the continuum phase velocity. Also, the
coefficient of $k^4$ comes with a negative sign, thus presumably leading to an
increase in group velocity, as well. 


\end{document}